\documentclass[aps]{revtex4}
\usepackage{epsf}
\usepackage{graphics}
\usepackage{amsmath}

\begin{document} 

\title{Entropy-induced smectic phases in rod-coil copolymers}

\author{Dominik D\"uchs$^{1}$ and D. E. Sullivan$^{2}$} 

\affiliation{$^{1}$Fakult\"at f\"ur Physik, Universit\"at Bielefeld,
Universit\"atsstr. 25, 33615 Bielefeld, Germany \\ 
$^{2}$Department of Physics and Guelph-Waterloo Physics Institute, University of Guelph,
Guelph, Ontario, Canada N1G 2W1}  
 
\begin{abstract}

We present a self-consistent field theory (SCFT) of semiflexible (wormlike) diblock copolymers,
each consisting of a rigid and a flexible part. The segments of the polymers are
otherwise identical, in particular with regard to their interactions, which are
taken to be of an Onsager excluded-volume type. The theory is developed in a
general three-dimensional form, as well as in a simpler one-dimensional
version. Using the latter, we demonstrate that the theory predicts the formation 
of a partial-bilayer smectic-A
phase in this system, as shown by profiles of the local density and orientational
distribution functions. The phase diagram of the system, which includes the 
isotropic and nematic phases, is obtained in terms of the mean density and 
rigid-rod fraction of each molecule. The nematic-smectic transition is found to be
second order. 
Since the smectic phase is induced solely by the difference in the
rigidities, the onset of smectic ordering is shown to be an entropic effect
and therefore does not have to rely on additional Flory-Huggins-type repulsive
interactions between unlike chain segments. These findings are compared with other 
recent SCFT studies of similar copolymer models and with computer simulations of 
several molecular models. 
\end{abstract}
\maketitle


\section{INTRODUCTION} \label{sec1}

Liquid crystals, being of major importance both on fundamental grounds and for industrial applications, have
long been of interest to researchers \cite{deGP,ggray}. Another class of materials of considerable interest are
melts of block copolymers, which have been shown in recent decades to exhibit
a multitude of phases of varying complexity \cite{Bates90}. This paper will be concerned with
liquid-crystalline copolymers, i.e., copolymers which exhibit phases commonly
associated with liquid crystals.

For the study of nonuniform polymer melts, a particular type of mean-field density-functional
approach traditionally known as self-consistent field theory (SCFT) has
proven to be a very powerful tool \cite{helfand,Nool,Fleer,Schmid98,Matsen3}. 
Aside from obvious 
determinants like the composition of the melt being studied, the models considered by 
SCFT approaches differ
mainly in two regards: the flexibility of the polymer chain and the interactions
between chain segments. Most commonly, polymers have been modelled 
as perfectly flexible Gaussian chains,
which assumes that there is no energy penalty for local bending. Alternatively, the
``wormlike'' chain model \cite{krp,fredr} does introduce such a bending penalty and
is thus the appropriate model for semiflexible polymers. Concerning the
interactions, most previous investigators have invoked Flory-Huggins-type
repulsive interactions between unlike chain segments in order to induce
microphase separation and hence the formation of various mesoscopic phases 
\cite{Matsen3,matsen1,netz}.

The aim of the present paper is to develop a modified SCFT suited to examining
the formation of liquid-crystalline phases in melts of diblock copolymers each consisting of a
rigid (``rod'') and a flexible (``coil'') part. Both parts are modelled as wormlike chains, but are
characterized by different rigidities. For the interactions
between any two chain segments, in this work we 
do not distinguish between the two parts, i.e., we assume that the same type of
interaction applies to any pair of chain segments. We take this interaction to
be the limit of the Onsager excluded-volume interaction for thin rods, which favors
local alignment of the chains. In the case of homogeneous chains, characterized
throughout by the same rigidity, the present model reduces to that applied by
Chen {\it et al}. \cite{chen,cui} to the study of nematic ordering
and the isotropic-nematic interface of semiflexible polymers. Here we show that the
generalized model with different rigidities for two parts of a polymer is able to 
account for the formation of lamellar smectic-A
phases. With this minimal model, the only
possible sources for the formation of such phases are entropic effects. Recently,
both smectic monolayer and bilayer
phases (traditionally denoted $A_1$ and $A_2$) have been found in an SCFT-based approach
which includes 
Flory-Huggins interactions and treats the 
rigid sections of the polymers
as perfectly rigid and perfectly aligned in the same orientation \cite{sem,matsen2}. In contrast, here we find 
only one type of smectic-A phase, which would most accurately be
classified as a partial bilayer phase (denoted $A_d$). In addition, as in Refs. \cite{chen,cui}, our model 
accounts for the disordered isotropic phase. 

Liquids of fairly {\it short}
rod-coil molecules have been examined recently by computer simulations 
\cite{sim0,sim1,sim2,sim3,allen,sim4}. Generally, these computer studies have shown that the 
addition of flexible segments to otherwise rigid molecules stabilizes the smectic-A
phase with respect to the nematic phase, 
consistent with experimental results for thermotropic non-polymeric liquid crystals
\cite{ggray}. This behavior is also shown by the present
model, although, in contrast to some studies \cite{sim1,sim2,allen}, we do not find 
that the nematic phase is entirely suppressed with respect to the isotropic and smectic phases.   
Our finding that the only stable smectic-A of the present model is the partial bilayer $A_d$ phase is in agreement
with a density-functional treatment by Holyst \cite{Hol} of rigid ``nail-shaped'' molecules,
and, somewhat more tentatively, with Monte Carlo studies by Mazars {\it et al.} \cite{sim2}
of a rod-coil model.

Experimental studies by J. T. Chen {\it et al}. \cite{chenJT} demonstrated 
both monolayer and bilayer smectic
ordering in rod-coil diblock copolymers consisting of a highly rigid
polyhexyl-isocyanate block joined to a flexible polystyrene coil. However, these
results were complicated by the simultaneous occurrence of tilted (or smectic-C)
ordering and crystallization of the rigid blocks. These two effects are not 
considered here, but will be examined in future work.

This paper is organized as follows. In Section II, a detailed description
of the theory is given. In Section III, numerical results are presented, showing
the local density and orientational order-parameter profiles in the smectic
phase as well as the phase diagram of the model in terms of mean density and rod fraction. 
We conclude with a summary and discussion in Section IV.
\\ \\

\section{THEORY} \label{sec2}

\subsection*{General 3-dimensional theory}
We consider a monodisperse melt of $n$ rod-coil diblock copolymers of total
contour length $L$, polymerization index $N$, and fixed segment length $a$ such that
$L=Na$, occupying a total volume $V$. A
fraction $\alpha$ of the total contour length of each copolymer is occupied by relatively 
rigid (rod) segments, and
the remaining fraction by more flexible (coil) segments. The total average monomer
density 
$\rho_0=Nn/V \equiv N\rho$ is fixed, whereas we shall not require the melt to fulfill a
$\it{local}$ incompressibility condition. In accordance with the wormlike chain model
for semiflexible chains \cite{krp,fredr,matsen1,netz,chen,cui}, which is applied here to both sections of the chains,
polymers will be treated as space curves ${\bf r}_{i}(t)$ characterized by dimensionless unit
tangent vectors ${\bf u}_{i}(t)$. Hence,
microscopic contour-averaged density operators may be defined as
\begin{eqnarray}
&&\hat\phi_{rigid}({\bf r},{\bf u})=\frac{1}{\rho}\sum_{i=1}^{n}\int_{0}^{\alpha}\mbox{d}t_i~
\delta\left({\bf r}-{\bf r}_{i}(t_{i})\right)
\delta\left({\bf u}-{\bf u}_{i}(t_{i})\right), \nonumber \\ 
&&\hat\phi_{flex}({\bf r},{\bf u})=\frac{1}{\rho}\sum_{i=1}^{n}\int_{\alpha}^{1}\mbox{d}t_i~ 
\delta\left({\bf r}-{\bf r}_{i}(t_{i})\right)
\delta\left({\bf u}-{\bf u}_{i}(t_{i})\right), \label{phihat}
\end{eqnarray}
which satisfy the normalization conditions
\begin{eqnarray}
&&\int\mbox{d}{\bf r}\mbox{ d}{\bf u}\mbox{
}\hat\phi_{rigid}({\bf r},{\bf u}) = \alpha V, \nonumber \\
&&\int\mbox{d}{\bf r}\mbox{ d}{\bf u}\mbox{
}\hat\phi_{flex}({\bf r},{\bf u}) = (1-\alpha) V.
\end{eqnarray}
Since these partial densities are only determined {\it a posteriori} and otherwise do
not enter the calculations, we shall instead use the total density,
\begin{equation}
\hat\phi({\bf r},{\bf u}) =
\hat\phi_{rigid}({\bf r},{\bf u}) + \hat\phi_{flex}({\bf r},{\bf u}).
\end{equation}
The interaction potential between any two segments, i.e., rod-rod, rod-coil, or
coil-coil, is taken to be the Onsager excluded-volume interaction, which in the
limit of very thin polymers ($L$,$a >>$ diameter $D$) reduces to \cite{cui} 
\begin{equation}
\label{Onsager}
v({\bf r_{1}},{\bf u_{1}};{\bf r_{2}},{\bf u_{2}})=L^{2}D\mbox{ }\delta({\bf r_{1}}-{\bf r_{2}})
\mbox{ }|{\bf u_{1}} \times {\bf u_{2}}|.
\end{equation}
The partition function in the canonical ensemble has the form
\begin{equation}
\label{Zfunc}
Z = \int{\cal D}_{n}\left\{\cdot\right\} \exp \left (- \rho C \int\mbox{d}{\bf r}
\int\mbox{d}{\bf u}\int\mbox{d}{\bf u}'\mbox{ }
\hat \phi({\bf r},{\bf u})\mbox{ }\hat \phi({\bf r},{\bf u}')\mbox{ }|{\bf u} \times
{\bf u}'| \right )~,
\end{equation}
where
\begin{equation}
C=L^{2} D \rho
\end{equation}
is proportional to the average polymer number density $\rho$. In Eq.(\ref{Zfunc}),
\begin{equation}
\int{\cal D}_{n}\left\{\cdot\right\}\equiv
\frac{1}{n!}\prod_{i=1}^{n}\int {\cal D} \left\{{\bf r}_{i},{\bf u}_{i}\right\} 
{\cal P}  \left\{{\bf r}_{i},{\bf u}_{i}[0,1]\right\},
\end{equation}
where
\begin{eqnarray}
\hspace{-1.2cm}
&&{\cal P} \left\{{\bf r}_{i},{\bf u}_{i}[s_{1},s_{2}]\right\} \propto \nonumber \\ 
\hspace{-1.2cm}
&&\prod_{t=s_{1}}^{s_{2}} \delta\left[{\bf u}_{i}(t)^{2}-1\right]
\delta\left[{\bf r}_{i}(t)-{\bf r}_{i}(s_{1})-L\int_{s_{1}}^{t}\mbox{d}s~{\bf u}_{i}(s)\right]
\exp\left[-\frac{1}{2N} \int_{s_{1}}^{s_{2}}\mbox{d}t'~
\kappa(t')\left|\frac{\mbox{d}{\bf u}_{i}}{\mbox{d}t'}\right|^{2}\right]
\end{eqnarray}
is the statistical weight of a given path and 
$\kappa(t)$
is a dimensionless bending modulus. According to the model considered here,
\begin{eqnarray}
\kappa(t) &=& \kappa_{rigid}~,~0 \le t < \alpha \nonumber \\
&=& \kappa_{flex}~,~\alpha < t \le 1~.
\end{eqnarray}

As is a standard procedure in self-consistent field theories, we now multiply
the partition function by $1 = \int {\cal D}\left\{\phi({\bf r},{\bf u})\right\} 
\delta(\phi({\bf r},{\bf u}) - \hat\phi({\bf r},{\bf u}))$, which allows us to replace the
operator $\hat\phi$ with the function $\phi$. Then, using the
exponential representation of the delta function, 
the  partition function can be rewritten as 
$Z\propto\int{\cal D}\left\{W\right\}\int{\cal D}\left\{\phi\right\}\exp(-{\cal F}[W,\phi])$ with 
\begin{eqnarray}
{\cal F} &=& \rho \Big[ C
\int \mbox{d}{\bf r}
\int\mbox{d}{\bf u}\int\mbox{d}{\bf u}'\mbox{ }
\phi({\bf r},{\bf u})\phi({\bf r},{\bf u}') |{\bf u} \times
{\bf u}'| \nonumber \\
&& - \int \mbox{d}{\bf r} \int\mbox{d}{\bf u}\mbox{ }
W({\bf r},{\bf u})\phi({\bf r},{\bf u})
\Big] 
- \ln \left(\frac{Q^{n}}{n!}\right), \label{F}
\end{eqnarray}
where
\begin{equation}
Q = \int{{\cal D}_{1}}\{\cdot\} \exp
\left(-\int_{0}^{1}\mbox{d}t\mbox{ }W\left({\bf r}(t),{\bf u}(t)\right)\right)
\end{equation}
is the single-chain partition function. The function $W({\bf r},{\bf u})$ 
is identified with the potential energy, or field, generated by all polymers in
the system as seen by a ``test'' polymer. 

Taking the saddle point of the free energy (\ref{F}) with respect to
$\phi({\bf r},{\bf u})$ and $W({\bf r},{\bf u})$, we obtain the mean-field equations:
\begin{subequations}
\begin{eqnarray}
W({\bf r},{\bf u}) &=& 2C\int\mbox{d}{\bf u}'\mbox{ }
\phi({\bf r},{\bf u}')|{\bf u}\times{\bf u}'|, \label{MFW} \\
\phi({\bf r},{\bf u}) &=& -\frac{V}{Q}\mbox{ }\frac{\delta Q}{\delta
W({\bf r},{\bf u})}. \label{MFphi}
\end{eqnarray}
\end{subequations}
In this approximation, the function $\phi({\bf r},{\bf u})$ equals the statistical
average $<\hat\phi({\bf r},{\bf u})>$ of the microscopic density. 
In order to solve these equations for the density,
we need to express $Q[W]$ in terms of the end-segment distribution function defined by
\begin{equation}
q({\bf r},{\bf u},t) = \int{\cal D}\{{\bf r}_{i},{\bf u}_{i}\}\mbox{ }{\cal P}\{{\bf r}_{i},{\bf u}_{i};[0,t]\}
\mbox{ }\delta({\bf r}-{\bf r}_{i}(t))\delta({\bf u}-{\bf u}_{i}(t))
\exp \left[ -\int_{0}^{t}\mbox{d}s~ W\left({\bf r}_{i}(s),{\bf u}_{i}(s)\right)\right].
\end{equation}
The function $q^{\dagger}({\bf r},{\bf u},t)$ is defined as the end-segment distribution
function starting from the opposite end of the polymer. Hence, 
\begin{equation}
Q = \int \mbox{d}{\bf u} \int \mbox{d}{\bf r}~ 
q({\bf r},{\bf u},t)q^{\dagger}({\bf r},{\bf u},t), 
\end{equation}
where the contour variable $t$ is arbitrary.
With the above definitions of $Q$, $q$, and $q^{\dagger}$, Eq.(\ref{MFphi}) yields 
for the local density:
\begin{equation}
\phi({\bf r},{\bf u}) = \frac{V}{Q}\int_{0}^{1}\mbox{d}t~
q({\bf r},{\bf u},t)q^{\dagger}({\bf r},{\bf u},t). \label{denseq} 
\end{equation}
The average rigid and flexible densities, $\phi_{rigid}({\bf r},{\bf u})$
and $\phi_{flex}({\bf r},{\bf u})$, are given by expressions analogous to (\ref{denseq}) on 
replacing the limits of integration over $t$ as in Eq.(\ref{phihat}).
The end-segment distribution functions, or propagators, satisfy
diffusion-like equations:
\begin{eqnarray}
\frac{\partial}{\partial t}q({\bf r},{\bf u},t)&=& 
\left[ -L{\bf u}\cdot\nabla_{{\bf r}} + \frac{1}{2\xi(t)}\nabla_{{\bf u}}^{2} -
W({\bf r},{\bf u})\right]~q({\bf r},{\bf u},t), \nonumber \\
\frac{\partial}{\partial t}q^{\dagger}({\bf r},{\bf u},t)&=& 
\left[ -L{\bf u}\cdot\nabla_{{\bf r}} - \frac{1}{2\xi(t)}\nabla_{{\bf u}}^{2} +
W({\bf r},{\bf u})\right]~q^{\dagger}({\bf r},{\bf u},t), \label{diffeq}
\end{eqnarray}
with initial conditions $q({\bf r},{\bf u},0)=1$ and
$q^{\dagger}({\bf r},{\bf u},1)=1$. Here we have defined the rigidity parameter
 $\xi(t) \equiv \kappa(t)/N$ depending on the chain contour variable $t$: $\xi(t)$
equals the persistence length of the corresponding chain section in units of the total contour
length $L$ \cite{matsen1}.
This is where the difference between the rigid and
flexible parts of the copolymer is accounted for.

To proceed with further analysis of the mean-field equations, we represent
the orientational (${\bf u}$) dependencies of the functions
$\phi, W, q$ and $q^{\dagger}$ using spherical-harmonic series:
\begin{eqnarray}
\phi({\bf r},{\bf u}) &=& \sum_{l,m} \phi_{lm}({\bf r})
Y_{l,m}({\bf u}), \nonumber \\
W({\bf r},{\bf u}) &=& \sum_{l,m} W_{lm}({\bf r}) Y_{l,m}({\bf u}),\nonumber \\
q({\bf r},{\bf u},t) &=& \sum_{l,m} q_{lm}({\bf r},t) Y_{l,m}({\bf u}), \nonumber \\
q^{\dagger}({\bf r},{\bf u},t) &=& \sum_{l,m} q^{\dagger}_{lm}({\bf r},t) Y_{l,m}({\bf u}).
\label{expansion}
\end{eqnarray}
Since these are all real functions, the expansion coefficients must obey the
following conditions:
\begin{equation}
\phi_{l,m}({\bf r}) = \phi^{*}_{l,-m}({\bf r})(-1)^m
\end{equation}
etc. Next, we expand the kernel $|{\bf u}\times{\bf u}'|$ by use of the addition theorem
for spherical harmonics \cite{cui},
\begin{equation}
|{\bf u} \times {\bf u}'| = \sum_{l,m} \frac{4\pi}{2l+1} d_{l}\mbox{ } Y_{l,m}({\bf u})
 Y^{*}_{l,m}({\bf u}') \label{intexp}
\end{equation}
with 
\begin{eqnarray}
d_{l} &=& 0~,~ l~\mbox{odd}, \nonumber \\
d_{0} &=& \frac{\pi}{4}~, \nonumber \\
d_{2k} &=& -\frac{\pi(4k+1)(2k)!(2k-2)!}{2^{4k+1}(k-1)!k!k!(k+1)!}~,~k=1,2,3,....
\end{eqnarray}
Inserting these formulae into the free energy (\ref{F}), the latter can be expressed as 
\begin{equation}
\label{Flm}
{\cal F} = \rho 
\sum_{l,m} \int\mbox{d}{\bf r}\Big[\frac{4\pi
C}{2l+1}d_{l} |\phi_{l,m}({\bf r})|^2
- \mbox{Re}\left(W_{l,m}({\bf r})\phi^*_{l,m}({\bf r})
\right)\Big] - \ln \frac{Q^{n}}{n!}~,
\end{equation}
where ``Re'' denotes real part.
The mean-field equations (\ref{MFW}) and (\ref{denseq}) now read
\begin{equation}
\label{Wlm}
W_{l,m}({\bf r}) = \frac{8\pi}{2l+1}d_{l} C \mbox{ } \phi_{l,m}({\bf r})~,
\end{equation}
\begin{eqnarray}
&&\phi_{l,m}({\bf r}) = 
  \frac{V}{Q}\int_{0}^{1} \mbox{d}t\sum_{l',m'}\sum_{l'',m''}
q^{\dagger}_{l',m'}({\bf r},t)q_{l'',m''}({\bf r},t)~ 
 \int d{\bf u} \mbox{ }Y_{l',m'}({\bf u}) Y_{l'',m''}({\bf u})  Y^{*}_{l,m}({\bf u})
\nonumber \\  \label{philm}
&&= \frac{V}{Q}\int_{0}^{1} \mbox{d}t \sum_{l',m'}\sum_{l'',m''}
q^{\dagger}_{l',m'}({\bf r},t)q_{l'',m''}({\bf r},t) 
 \sqrt{\frac{(2l''+1)(2l'+1)}{4\pi(2l+1)}} C^{l'',l',l}_{0,0,0}
C^{l'',l',l}_{m'',m',m}~,
\end{eqnarray}
with 
\begin{equation}
Q = \sum_{l,m} \int\mbox{d}{\bf r}\mbox{ } 
q_{l,m}({\bf r},t)q^{\dagger}_{l,-m}({\bf r},t)(-1)^{m}.
\end{equation}
The $C^{l'',l',l}_{m'',m',m}$ are Clebsch-Gordan
coefficients, and we have used a result for the integral of three spherical harmonics 
\cite{Gray}. The corresponding projections $\phi_{rigid,l,m}$ and $\phi_{flex,l,m}$ are obtained likewise by
changing the limits of the line integral in Eq.(\ref{philm}). In terms of the projections $q_{l,m}
({\bf r},t)$, the diffusion-like equation (\ref{diffeq}) yields the following coupled
set of equations:
\begin{eqnarray}
\hspace{-1cm}
\frac{\partial}{\partial t}q_{l,m}({\bf r},t) = &-&
L\sum_{l',m'}\sqrt{\frac{2l'+1}{2l+1}}\Big[ C^{1,l',l}_{0,0,0}
\Big( - \frac{1}{\sqrt{2}}\left(C^{1,l',l}_{1,m',m}+C^{1,l',l}_{-1,m',m}\right)\frac{\partial}{\partial
x} \nonumber \\
\hspace{-1cm}
&+& \frac{i}{\sqrt{2}}\left(C^{1,l',l}_{1,m',m}-C^{1,l',l}_{-1,m',m}\right)\frac{\partial}{\partial
y} +C^{1,l',l}_{0,m',m} \frac{\partial}{\partial z} \Big)\Big]q_{l',m'}({\bf r},t)
\nonumber \\
\hspace{-1cm}
&-& \frac{1}{2\xi(t)} l(l+1) q_{l,m}({\bf r},t) \nonumber \\ 
\hspace{-1cm}
&-& \sum_{l',l'',m',m''}\sqrt{\frac{(2l''+1)(2l'+1)}{4\pi(2l+1)}} W_{l',m'}({\bf r})q_{l'',m''}({\bf r},t)
C^{l'',l',l}_{0,0,0}C^{l'',l',l}_{m'',m',m} \label{diffdiff}
\end{eqnarray}
with initial conditions 
\begin{eqnarray}
q_{0,0}({\bf r},0) &=& \sqrt{4\pi}~, \nonumber \\
q_{l,m}({\bf r},0) &=& 0~, \mbox{ otherwise}.
\end{eqnarray}
A similar equation applies to $q^{\dagger}({\bf r},{\bf u},t)$.

\subsection*{1-dimensional theory}
For simple applications of the general theory
presented above, we now specialize to situations where the densities vary in
only one spatial
dimension, which for convenience is chosen to be the $z$ direction. 
Furthermore, we will restrict analysis to phases that exhibit
no azimuthal orientation dependence about the $z$-axis, thereby excluding the possibility of 
smectic-C phases \cite{matsen2,sim3}. Then the only nonzero projections of
any angular functions are those with $m=0$, so that we will subsequently drop the $m$
indices, denoting $\phi_{l}=\phi_{l,m=0}$, etc.
Another consequence is that now all spherical-harmonic expansion coefficients are real.
The free energy Eq.(\ref{Flm}) becomes
\begin{equation}
{\cal F} = \rho A 
\sum_{l}\int\mbox{d}z\mbox{ } \left[\frac{4\pi
C}{2l+1}d_{l}\phi_{l}^{2}(z) 
- W_{l}(z)\phi_{l}(z)\right] - \ln \frac{Q^{n}}{n!}~,
\end{equation}
where $A$ is the cross-sectional area of the system in the $x$ and $y$ directions.
The mean-field equations (\ref{Wlm}) and (\ref{philm}) are:
\begin{subequations}
\begin{eqnarray}
W_{l}(z) &=& \frac{8\pi}{2l+1}d_{l} C \mbox{ } \phi_{l}(z), \label{mf1}\\
\phi_{l}(z) &=& \frac{V}{Q} \sum_{l',l''} \int_{0}^{1}\mbox{d}t~ 
q^{\dagger}_{l'}(z,t)q_{l''}(z,t) 
\sqrt{\frac{(2l''+1)(2l'+1)}{4\pi(2l+1)}} \left(C^{l'',l',l}_{0,0,0}\right)^{2} \label{mf2}
\end{eqnarray}
\end{subequations}
with
\begin{equation}
Q = A \sum_{l} \int \mbox{d}z~ 
q^{\dagger}_{l}(z,t)q_{l}(z,t)~.
\end{equation}
Finally, the diffusion-like equation (\ref{diffdiff}) becomes
\begin{eqnarray}
\frac{\partial}{\partial t}q_{l}(z,t) = &-& L
\sum_{l'}\sqrt{\frac{2l'+1}{2l+1}}\left(C^{1,l',l}_{0,0,0}\right)^{2}
\frac{\partial}{\partial z} q_{l'}(z,t)
- \frac{1}{2\xi(t)} l(l+1) q_{l}(z,t) \nonumber \\ 
&-& \sum_{l',l''}\sqrt{\frac{(2l''+1)(2l'+1)}{4\pi(2l+1)}} \left(C^{l'',l',l}_{0,0,0}\right)^{2}
W_{l'}(z)q_{l''}(z,t) \label{diff}
\end{eqnarray}
with initial conditions
\begin{eqnarray}
q_{0}(z,0) &=& \sqrt{4\pi}~, \nonumber \\
q_{l}(z,0) &=& 0 \mbox{ , } l>0.
\end{eqnarray}

In the following, we shall truncate the interaction expansion (\ref{intexp}) after $l=2$,
(equivalent to a ``Maier-Saupe'' interaction), retaining only the terms with
coefficients $d_{0}=\frac{\pi}{4}$ and
$d_{2}=-\frac{5\pi}{32}$. 

We shall now briefly discuss the
computational methods used in solving the theory. 
The fields and densities are determined self-consistently according to
Eqs.(\ref{mf1}), (\ref{mf2}) and (\ref{diff}) using a fixed-point iteration algorithm with
variable mixing parameters for successive iterations. An iteration consists of:
(a) given the functions $W_l(z)$, solving Eq. (\ref{diff}) and its counterpart for 
$q_l^{\dagger}$; (b) 
calculating the set of functions $\phi_l(z)$ from
Eq. (\ref{mf2}); (c) obtaining a new set of functions $W'_l(z)$ from Eq. (\ref{mf1}); (d) mixing 
$W'_l(z)$ and $W_l(z)$ 
according to a fixed-point algorithm, which yields a new $W_l(z)$ for the
next iteration. 

Concerning step (a), solutions of the
diffusion-like equation (\ref{diff}) were discretized in time and space according to a Forward
Time Centered Space (FTCS) scheme \cite{nr}, which is explicit and
straightforward to implement. 
This is a reliable discretization scheme for solving
partial differential equations although its error is only first order in the contour discretization
$dt$. For the corresponding diffusion equation 
in the case of Gaussian chains,
a number of more sophisticated schemes such as
Crank-Nicholson and DuFort-Frankel \cite{CNDF} are available to improve on
stability and accuracy (the error becomes second order in $dt$ in both cases),
and thus to lower the contour resolution needed for a 
given desired level of accuracy, a key factor in computing times. Due to the
coupled nature of Eq.(\ref{diff}), however, it seems daunting to apply
similarly efficient discretization schemes to the wormlike chain model. 

Concerning step (d), the fixed-point iteration algorithm used in finding self-consistent solutions
of Eqs.(\ref{mf1}) and (\ref{mf2}) is a tried and true method. Here
viable alternatives do exist, however, first and foremost being Newton-Raphson-type
algorithms like Broyden's method (e.g. \cite{broyden}). In the present work, we have 
restricted ourselves
to using the fixed-point algorithm, since our primary aim was to develop the
framework presented above and illustrate it with only a few, demonstrative results.
					
In the present work, calculations were performed on a 1-dimensional grid with
periodic boundary conditions, a spatial discretization of $dz=0.02$, and a
contour discretization of $dt=1/1500$. The spherical-harmonic expansions of the 
propagators and
densities were truncated after $l=12$. With these parameters, ${\cal F}/V$ can be
determined to within a numerical inaccuracy of less than $1 \%$, requiring up to 500
iterations, or 3-4 hours on a Pentium II processor. Note that although 
the interaction kernel (\ref{intexp}) is truncated after $l=2$, we cannot truncate the
propagators at 
the same value of $l$, since the coupling in (\ref{diff}) renders these
higher-order projections nonzero.

Once Eqs.(\ref{mf1}), (\ref{mf2}) and (\ref{diff}) have been solved self-consistently,
the free energy can be rewritten as
\begin{equation}
{\cal F} = -4\pi C\rho A 
\sum_{l} \frac{d_{l}}{2l+1}\int\mbox{d}z\mbox{ }\phi_{l}^{2}(z) 
- \ln \frac{Q^{n}}{n!}.
\end{equation}
For determining coexistence regions of the first-order isotropic-nematic
transition, it is necessary to perform double-tangent constructions on curves of the 
free energy per volume ${\cal F}/V$.  
Omitting linear terms in $C$ (which yield constants when the derivative 
$\frac{\partial}{\partial C}$ is taken), we obtain
\begin{equation}
\frac{{\cal F}}{V} \propto \frac{{\cal F}C}{n} 
 = -4\pi C^{2} \left(\frac{A}{V}\right) 
\sum_{l} \frac{d_{l}}{2l+1}\int\mbox{d}z\mbox{ }\phi_{l}^{2}(z) 
- C \ln \frac{Q}{V} + C \ln C,
\end{equation}
where we have used Stirling's approximation for the factorial. In a smectic phase,
all projections of the local densities and propagators are taken to be periodic functions
of $z$ with period (or layer spacing) $D$. All integrations over $z$ can then be
evaluated as
\begin{equation}
\frac{A}{V} \int\mbox{d}z = \frac{1}{D} \int_{0}^{D} \mbox{d}z~.
\end{equation} 
The location of the second-order nematic-smectic transition
(cf. section III) is determined by the behavior of the smectic order parameter
defined as:
\begin{equation}
O_{sm} = \left[\frac{1}{D}\int_{0}^{D} \mbox{d}z~ \left( [\sqrt{4\pi}\phi_{rigid,0}-\alpha]^{2} +
[\sqrt{4\pi}\phi_{flex,0}-(1-\alpha)]^{2} \right)\right]^{1/2}.
\label{Oformula}
\end{equation}
The parameter $O_{sm}$ vanishes in the isotropic and nematic phases, while it is
nonzero in the smectic phase. For a ``perfect'' smectic phase in which the profiles
$\sqrt{4\pi}\phi_{rigid,0}$ and $\sqrt{4\pi}\phi_{flex,0}$ have rectangular shapes
of widths $\alpha D$ and $(1-\alpha)D$, respectively, $O_{sm}$ has the value
$\sqrt{2 \alpha (1-\alpha)}$.
The equilibrium period is that value of $D$ which minimizes the free energy per volume.

\section{RESULTS} \label{sec3}

The results presented here apply to a fluid  of polymers with
$\xi_{rigid}$ = 10, and $\xi_{flex}$ = 0.1. These values more or
less represent the feasible bounds on $\xi$ for
performing numerical calculations. To begin with, we shall examine a system
with $\alpha = 2/3$. 

\begin{figure}[h]
\center
\begin{minipage}{5in}
\epsfxsize= 5in \epsfbox{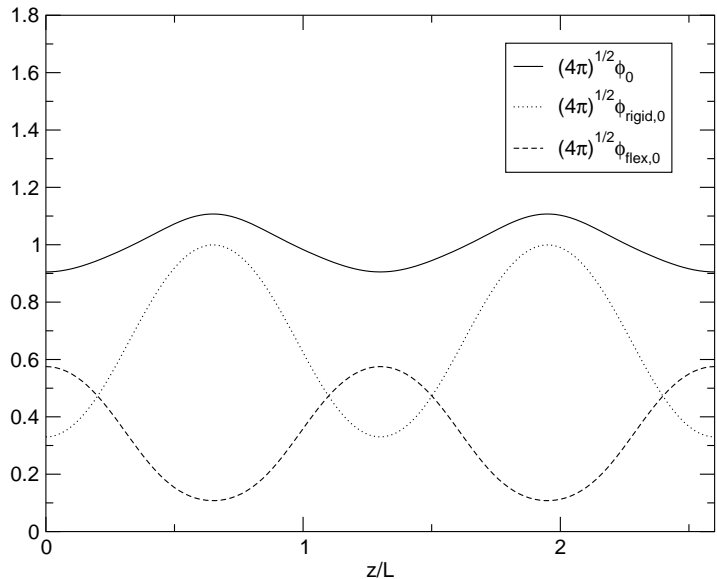}
\end{minipage}
\renewcommand
\baselinestretch{1.}
\caption{Density profiles of a smectic configuration with $\alpha=2/3$, 
$\xi_{rigid}=10$, $\xi_{flex}=0.1$, $C=20$, and period $1.3L$. The solid,
dotted, and dashed lines correspond to $\sqrt{4\pi}\phi_{0}$,
$\sqrt{4\pi}\phi_{rigid,0}$, and $\sqrt{4\pi}\phi_{flex,0}$, respectively.}
\renewcommand
\baselinestretch{1.5}
\label{smec1}
\end{figure}

Fig. \ref{smec1} shows the rigid and flexible density profiles (i.e., zeroth-order
projections $\phi_{\beta,l=0}(z)$, $\beta \in \{\mbox{rigid,flex}\}$) 
as well as the total density profile $\phi_0(z)$ in a smectic
configuration at $C=20$, just above the nematic-smectic transition. We see
a clearly defined region of predominantly rigid chain segments and a less
pronounced region where the flexible parts are dominant. The individual density
variations are nearly pure sine functions, as expected close to a second-order
transition. 
The total density in this lamellar structure exhibits maxima in
the rigid regions and minima in the flexible regions. This indicates that
the rigid segments pack more efficiently than flexible ones due to their greater
susceptibility  
to local orientational alignment.

\begin{figure}[h]
\center
\begin{minipage}{5in}
\epsfxsize= 5in \epsfbox{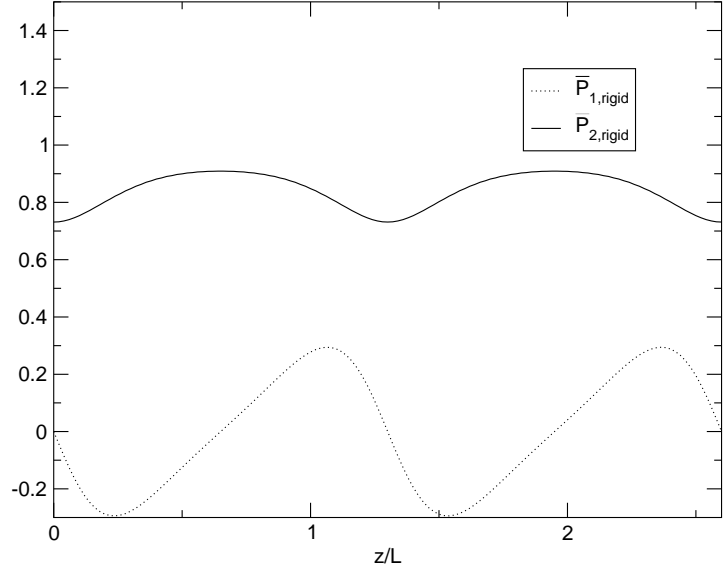}
\end{minipage}
\renewcommand
\baselinestretch{1.}
\caption{
Orientational order parameters $\bar P_{1,rigid}$ (dotted line) and $\bar P_{2,rigid}$ (solid line) for the same
case as in Fig. 1.
}
\renewcommand
\baselinestretch{1.5}
\label{Pr20}
\end{figure}

\begin{figure}[h]
\center
\begin{minipage}{5in}
\epsfxsize= 5in \epsfbox{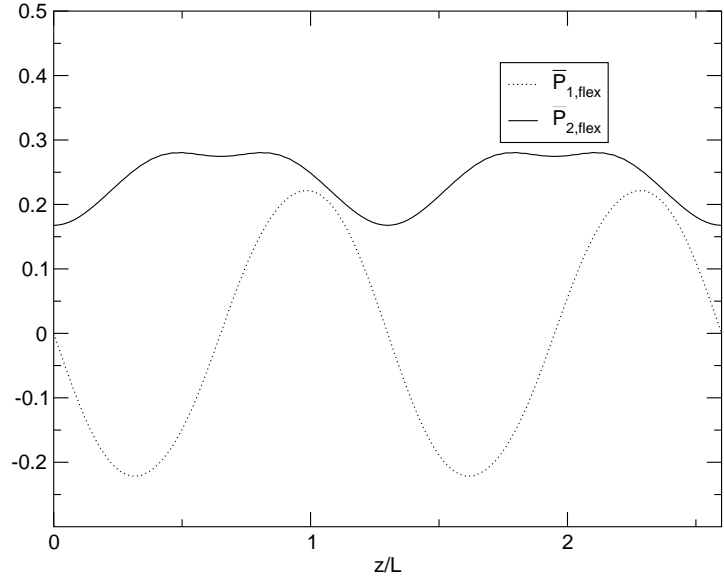}
\end{minipage}
\renewcommand
\baselinestretch{1.}
\caption{
Order parameters $\bar P_{1,flex}$ (dotted line) and $\bar P_{2,flex}$ (solid line) for the same
case as in Fig. 1. 
}
\renewcommand
\baselinestretch{1.5}
\label{Pf20}
\end{figure}

More insight can be gained from the orientational projections of the 
densities, $\phi_{\beta,l}(z)$. Here we express these in terms of the $z$-dependent
order parameters defined as, for each degree $l$, 

\begin{equation}
\bar P_{l,\beta}(z) = \frac{1}{\sqrt{2l+1}}~\frac{\phi_{\beta,l}(z)}{\phi_{\beta,0}(z)}~,
\end{equation}
equivalent to the average of the $l^{th}$ Legendre polynomial $P_l(\cos\theta)$
{\it per segment} of species $\beta$, where $\theta$ is the angle between a segment
axis and the $z$ axis. The lowest-order functions $\bar P_{1,\beta}(z)$ and
$\bar P_{2,\beta}(z)$ corresponding to the configuration of Fig. \ref{smec1}
are shown in
Figs. \ref{Pr20} and \ref{Pf20}. The second-order distribution $\bar P_{2,\beta}$
characterizes the overall degree of orientational order of species $\beta$, while 
$\bar P_{1,\beta}$ indicates the average spatial direction in which segments $\beta$
are oriented. The behavior of the functions $\bar P_{1,\beta}(z)$ 
shows that both the rigid and flexible regions are divided in the middle
into domains of positive and negative orientation along $z$, in the manner of a
bilayer. 
We see that both $\bar P_{2,rigid}$ and $\bar P_{2,flex}$
exhibit maxima (minima) in the regions of high (low) rod density
(cf. Fig. \ref{smec1}). The minimum in the flexible coil distribution 
$\bar P_{2,flex}$ is more pronounced, while its maximal region is a slightly
modulated plateau within the domain of high rigid density, suggesting 
that coil segments are oriented by interactions with neighboring rigid segments. Overall,
the coils exhibit a much weaker degree of orientational ordering than the rods.

\begin{figure}[h]
\center
\begin{minipage}{5in}
\epsfxsize= 5in \epsfbox{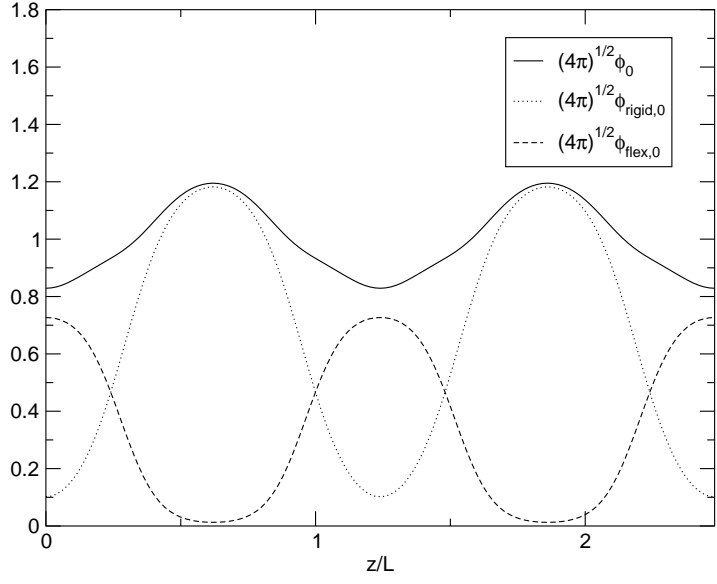}
\end{minipage}
\renewcommand
\baselinestretch{1.}
\caption{Density profiles as in Fig. 1 but for $\alpha=2/3$, 
$C=30$, and period $1.24L$.}
\renewcommand
\baselinestretch{1.5}
\label{C30}
\end{figure}

\begin{figure}[h]
\center
\begin{minipage}{5in}
\epsfxsize= 5in \epsfbox{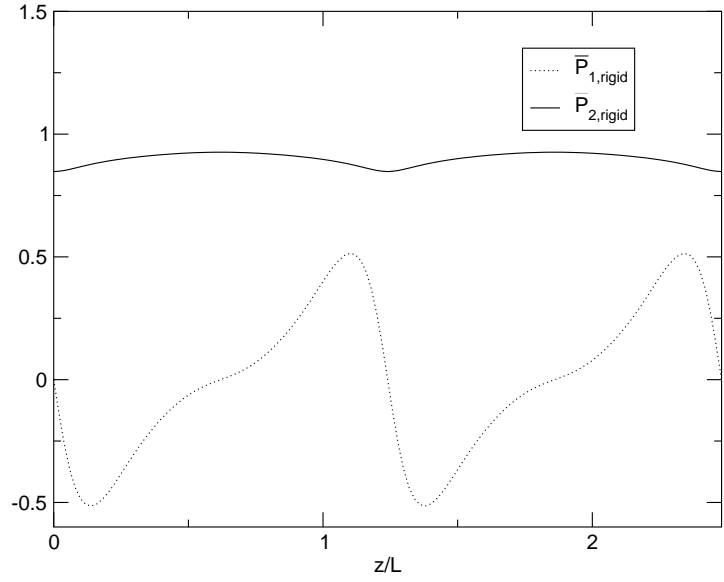}
\end{minipage}
\renewcommand
\baselinestretch{1.}
\caption{
Orientational order parameters of rigid segments for the same case as in Fig. 4. 
}
\renewcommand
\baselinestretch{1.5}
\label{Pr30}
\end{figure}

\begin{figure}[h]
\center
\begin{minipage}{5in}
\epsfxsize= 5in \epsfbox{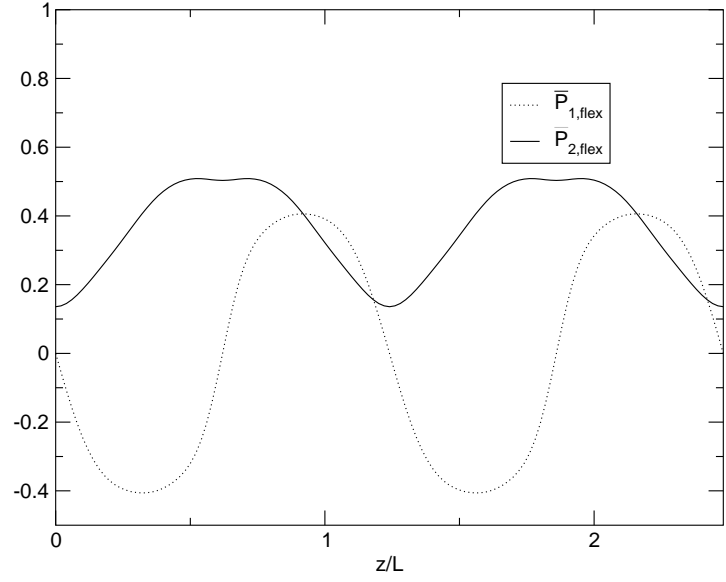}
\end{minipage}
\renewcommand
\baselinestretch{1.}
\caption{
Orientational order parameters of flexible segments for the same case as in Fig. 4.
}
\renewcommand
\baselinestretch{1.5}
\label{Pf30}
\end{figure}

The configuration shown in Figs. \ref{smec1}, \ref{Pr20} and \ref{Pf20} has an optimal period  
(i.e., yielding the minimum free energy per volume) of $D = 1.30L$ for the chosen parameters
$\alpha= 2/3$ and $C = 20$. Keeping the same $\alpha$ but increasing $C$ to $30$ results
in a decrease of the optimal period to $D=1.24$. This is accompanied by 
stronger segregation of the rod and coil regions and less purely
sinusoidal behavior of the densities, as shown in Fig. \ref{C30}. The corresponding
orientational order parameters are given in Figs. \ref{Pr30} and \ref{Pf30}, and are
qualitatively similar to those for $C=20$. We have found 
that the optimal periods range from $1.1L$ for $\alpha=0.3$ to $1.3L$ for
$\alpha=0.75$ for values of $C$ close to the $N - A_{d}$ transition. Smectic phases
with values of $\alpha$ outside this interval could not be examined due to
numerical difficulties. Attempts to generate solutions corresponding to
monolayer smectic structures having $D \leq L$ always converged to a uniform
(isotropic or nematic) phase. 

Fig. \ref{phdiagr10} shows the phase diagram in the $C-\alpha$ plane for rod-coil polymers with
$\xi_{rigid}=10$ and $\xi_{flex}=0.1$. At low mean densities $C$, the steric
interactions do 
not suffice to generate an ordered phase: the system is isotropic. Upon increasing
$C$, we encounter a first-order isotropic-nematic ($I-N$) transition. The two branches
delineating the coexistence region in
the diagram were determined by constructing tangents to the ${\cal F}/V$ graphs for
the isotropic and nematic solutions with a spline interpolation, which amounts
to calculating the values of $C$ at equal chemical potentials and pressures.
The lower $\alpha$, the higher is the value of $C$ at which the $I-N$ transition occurs: this
transition is driven primarily by the tendency of the rigid-rod segments to align and thus
requires higher densities if the proportion of the rods is lowered. 
For $\alpha=1$ we can compare our data with that of Chen \cite{chen}, who
uses the full kernel (\ref{intexp}) and obtains an isotropic density of 
$C_{iso}=4.18$ and a nematic density of $C_{nem}=5.33$. If we calculate
this transition retaining the kernel up to $l=12$, we obtain good agreement:
$C_{iso}=4.16$ and $C_{nem}=5.29$. With the truncated kernel ($l\le2$)
used in the majority of this work, the coexistence region becomes more narrow:
$C_{iso}=4.89$ and $C_{nem}=5.35$.

\begin{figure}[h]
\center
\begin{minipage}{5in}
\epsfxsize= 5in \epsfbox{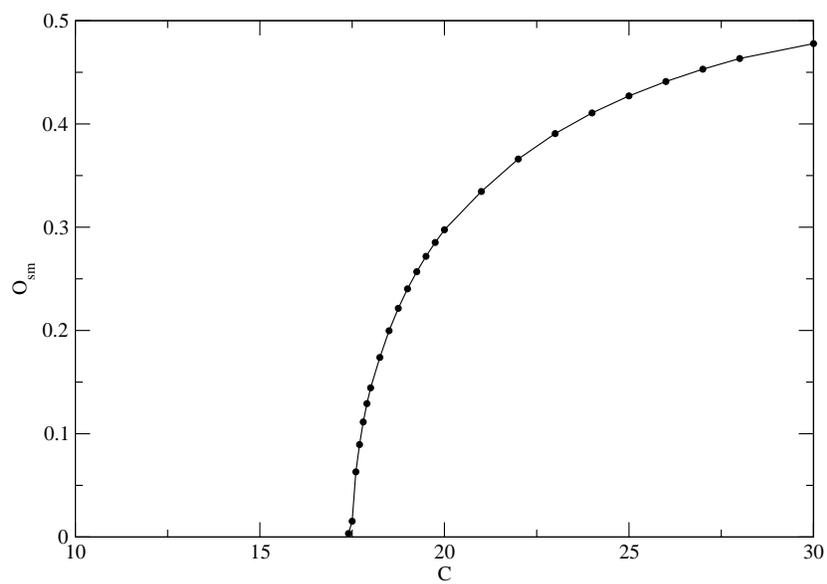}
\end{minipage}
\renewcommand
\baselinestretch{1.}
\caption{Smectic order parameter vs. $C$ for $\alpha=2/3.$}
\renewcommand
\baselinestretch{1.5}
\label{Osmec}
\end{figure}

A second transition, from the nematic to the smectic ($A_d$) phase, occurs at higher values of
$C$. This transition is second-order within numerical uncertainties, as is
indicated by the behavior of the smectic order parameter $O_{sm}$, Eq. (\ref{Oformula}).
Fig. \ref{Osmec} shows that $O_{sm}$ starts to grow from zero at a well-defined
critical value of the density $C$. There is a minimum of the $N-A_d$ transition
density at roughly $\alpha=0.55$.
On moving toward both lower and higher values of $\alpha$, the
critical values of $C$ increase sharply until numerical difficulties prevent us from
extending the graph further. Obviously, a well-balanced proportion of rod-like
vs. coil-like segments facilitates forming a smectic phase: the former stabilize
the rod-dominated portion of the density profile (the nematic microdomain, as
it 
were), and the latter, the coil-dominated portion. Deviations from the optimal
$\alpha$ to higher values of
$\alpha$ decrease the entropic advantage of the coils, which has to be compensated
for by an increased density. Lower values of $\alpha$, on the other hand, destabilize the
nematic microdomain ordering of the rods. This is why, for even lower values of $\alpha$,
the $I-N$ and $N-A_d$ phase boundaries approach each other until intersecting at $\alpha=0.32$.
Below this value of $\alpha$, the $I-N$ 
transition is preempted by a first-order $I-A_d$ transition. For $\alpha=0.3$, we
find a coexistence region between $C_{iso}=29.3$ and $C_{smec}=31.4$.
Unfortunately, numerical problems prevented us from examining even lower values of
$\alpha$: it thus remains an open question by how much the $I-A_d$ transition 
deviates from the preempted $I-N$ transition.

\section{SUMMARY AND CONCLUSIONS} \label{sec4}

We have presented a self-consistent field theory for semiflexible copolymers in a
general three-dimensional, as well as a one-dimensional form without azimuthal orientational
dependence. Using this framework, numerical calculations have been performed
to investigate the occurrence of a smectic-A phase in systems of rod-coil copolymers where
the two parts of each molecule differ only in their rigidities.

\begin{figure}[h]
\center
\begin{minipage}{5in}
\epsfxsize= 5in \epsfbox{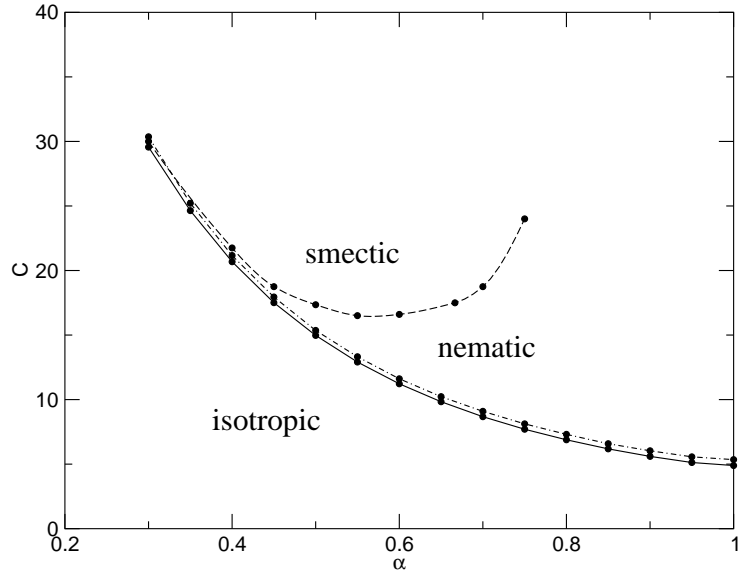}
\end{minipage}
\renewcommand
\baselinestretch{1.}
\caption{Phase diagram for rod-coil copolymers with $\xi_{rigid}=10$, $\xi_{flex}=0.1$. The
isotropic and nematic phases are separated by a narrow coexistence region bounded by the
solid and dash-dotted curves. The boundary between smectic and nematic phases is second
order, indicated by the dashed curve.
}
\renewcommand
\baselinestretch{1.5}
\label{phdiagr10}
\end{figure}

A phase diagram was established for $\xi_{rigid}=10$ and $\xi_{flex}=0.1$,
which correspond to very stiff rods and very flexible coils. At low densities $C$, an
isotropic phase is present, which becomes nematic for larger values of $C$ via a
first-order transition. At even larger values of $C$, a partial bilayer smectic-A phase forms,
whose period increases with increasing rod fraction $\alpha$. The latter effect,
while small in the examined parameter range, indicates that the coils can
overlap with other coil segments as well as with the rods, resulting in 
interdigitation of the layers. This feature is also suggested by the rather
diffuse, sinusoidal variation of the local density profiles $\phi_{\beta,0}$. 
The shape of the smectic phase boundary in Fig. \ref{phdiagr10}, exhibiting 
a minimum mean density $C$ near a rod fraction $\alpha = 0.55$, is plausible and consistent
with experimental deductions \cite{sim3}. (See also \cite{mul}.)  
Preliminary calculations indicate that lowering the rod rigidity $\xi_{rigid}$
shifts all transitions to higher densities $C$.

In the present work, only anisotropic excluded-volume interactions between chain segments,
assumed to be the same for the rigid and flexible portions of each molecule, 
are taken into account. To generate smectic phases, we have shown 
that it is sufficient to distinguish rigid and flexible parts of a polymer by means of
their rigidities or reduced persistence lengths $\xi$, and thus have demonstrated that lamellar
ordering is a purely entropic phenomenon. This is consistent with the findings of 
computer simulations \cite{sim0,sim2,sim3,allen,sim4} and a density-functional treatment
\cite{Hol} of diblock models employing only anisotropic repulsive intermolecular forces.
In contrast, the smectic ordering found in the SCFT treatments of Refs. \cite{sem,matsen2}
is driven primarily by {\it isotropic} Flory-Huggins interactions between unlike chain
segments, which may well be present in more realistic models. Unlike here,
the latter works predicted the formation of both monolayer and bilayer smectic phases,
as well as strong segregation between rigid and flexible domains,
which we attribute both to the quite different nature of the interactions adopted and
to the allowance of local compressibility in the present theory: as conjectured in Ref. 
\cite{matsen2}, compressibility effects ``could  dramatically stabilize the bilayer phase''. 

We expect that in a more accurate treatment of excluded-volume effects,
the density $C$ at the $N-A_d$ transition will not increase so sharply with 
increasing $\alpha$ as indicated in Fig. \ref{phdiagr10}. Note that the Onsager
excluded-volume interaction Eq. (\ref{Onsager}) is taken to be spatially local, 
and hence by itself does not generate smectic phases in the absence of differing
chain rigidities.  A more realistic non-local treatment of the repulsive interactions
between chain segments should generate a smectic-A phase even in the case of  
homogenous semiflexible polymers \cite{Sear,bladon,schoot,dogic}.

Future extensions of the theory presented here will attempt to account for the non-local
interactions mentioned in the preceding paragraph as well as for
smectic-C phases, crystallization of the rods, and non-lamellar morphologies, all of which are
expected to occur at larger values of $C$ and more extreme values of the rod fraction $\alpha$.

We thank F. Schmid and M. Matsen for fruitful discussions. This work was 
supported through grants from the Deutsche Forschungsgemeinschaft (DFG) and 
the Natural Sciences and Engineering Research Council (Canada) (NSERC).


\end{document}